# Shashthosheba: Dissecting Perception of Bangladeshi People towards Telemedicine Apps through the Lens of Features of the Apps

Waqar Hassan Khan
1505107.whk@ugrad.cse.buet.ac.bd
Bangladesh University of Engineering
and Technology
Dhaka, Bangladesh

Md Al Imran
1505100.mai@ugrad.cse.buet.ac.bd
Bangladesh University of Engineering
and Technology
Dhaka, Bangladesh

Ahmed Nafis Fuad
1505113.anf@ugrad.cse.buet.ac.bd
Bangladesh University of Engineering
and Technology
Dhaka, Bangladesh

Mohammed Latif Siddiq
msiddiq3@nd.edu
University of Notre Dame
Notre Dame, Indiana, USA

A. B. M. Alim Al Islam
alim_razi@cse.buet.ac.bd
Bangladesh University of Engineering
and Technology
Dhaka, Bangladesh

## ABSTRACT

Bangladesh, a developing country with a large and dense population, has recently seen significant economic as well as technological developments. The growth of technology has resulted in a dramatic increase in the number of smartphone users in Bangladesh, and as such, mobile apps have become an increasingly important part of peoples' life, even encompassing healthcare services. However, the apps used in healthcare (telemedicine to be specific) in Bangladesh are yet to be studied *from the perspective of their features* as per the voices of the users as well as service providers. Therefore, in this study, we focus on the features of the telemedicine apps used in Bangladesh. First, we evaluated the present status of existing telemedicine apps in Bangladesh, as well as their benefits and drawbacks in the context of HCI. We analyzed publicly accessible reviews of several Bangladeshi telemedicine apps (N = 14) to evaluate the user impressions. Additionally, to ascertain the public opinion of these apps, we performed a survey in which the patients (N = 87) participated willingly. Our analysis of the collected opinions reveals what users experience, what they appreciate, and what they are concerned about when they use telemedicine apps. Additionally, our study demonstrates what users expect from telemedicine apps, independent of their past experience. Finally, we explore how to address the issues we discovered and how telemedicine may be used to effectively offer healthcare services throughout the country. To the best of our knowledge, this study is the first to analyze the perception of the people of Bangladesh towards telemedicine apps from the perspective of features of the apps.

## 1 INTRODUCTION

Bangladesh is a developing country situated in South Asia. As of January 2022, it is the world's eighth-most populated nation with a very high population density. With rapid economic growth, [15] as well as technology adoption nationwide [35, 44], the usage of the Internet and smartphones are on the rise. More than 120 million people are now connected to the Internet [10] in Bangladesh.

The growing use of the Internet and smartphones has resulted in a rise in various smartphone apps. Ride-sharing apps, online food delivery apps, online shopping apps, mobile financial system apps, etc., are now becoming an integral part of people's lives in the country. These apps are not only simplifying citizen's life but also enabling a substantial part of the society to make a living [45, 57]. Especially during the lockdown due to COVID-19, when institutions, shopping malls, and restaurants were closed, these apps became nearly unavoidable in maintaining social distancing and preventing the spread of COVID-19 [61]. Like the other examples, Telemedicine apps were also crucial in assuring healthcare during the pandemic. Such apps remain important nowadays and for the future, considering the far-reaching effects ahead. According to WHO (World Health Organization), telemedicine has been defined as follows: *The delivery of health care services, where distance is a critical factor, by all health care professionals using information and communication technologies for the exchange of valid information for diagnosis, treatment and prevention of disease and injuries, research and evaluation, and for the continuing education of health care providers, all in the interests of advancing the health of individuals and their communities* [49].

The people of Bangladesh, in general, are accustomed to physically consulting doctors. Additionally, the majority of highly qualified and experienced doctors reside in cities, particularly in the capital city Dhaka [4, 32]. As a result, it is challenging for rural residents, who are around 62% of the overall population [16], to consult with experienced doctors. As an unavoidable consequence due to the unavailability and high cost of experienced doctors, people in rural areas rely heavily on medicine shopkeepers [62]. On the other hand, even though qualified doctors are available in cities, visiting hospitals during a pandemic have become highly dangerous.

Numerous doctors tested positive after coming into contact with COVID-19 positive patients unknowingly, and there were several cases of deaths of doctors in Dhaka, the capital of Bangladesh.

Considering all these aspects from both sides, i.e., patients and doctors, telemedicine apps have increased more than ever in Bangladesh. Several telemedicine apps are now available for Bangladeshi citizens on Google Play[1]. Doctor Dekhao, DocTime, Digital Hospital, My Health, and Sebaghar are just a few of the most popular and highly-rated apps in this regard. These telemedicine apps enable patients to consult with doctors without meeting them in person. These apps are compatible with both smartphones and tablets. The majority of apps on the market include standard functionalities such as searching for doctors on the platform, viewing their schedules, scheduling appointments, consulting with them via video or audio calls, receiving digital prescriptions, uploading reports, booking diagnostic tests, etc. Such functionalities or features are the core foundation of the service provided by the apps. Therefore, it is of utmost significance to have an in-depth study of the features of telemedicine apps. However, to the best of our knowledge, such a study has a core focus on the features of telemedicine apps for Bangladesh is yet to be done in the literature. This happens as the existing research studies mainly focus on the telemedicine apps overall as a service [23, 48, 53, 56, 65] from a course granularity rather than going deep towards the individual features of the apps at a fine granularity.

## 1.1 Research Questions and Goal of This Study

While focusing on features of telemedicine apps from the perspectives of Bangladesh, we have identified the following research questions,

**RQ1:** *What do users think about features of the telemedicine apps currently in operation in Bangladesh?*
This question covers the features of current telemedicine apps, the issues users encountered, and the user interface/user experience (UI/UX) of the apps. To address this research question, we gathered and carefully examined reviews from Google Play.

**RQ2:** *What features do the users in Bangladesh want in telemedicine apps?*
Users often express a need for specific features in apps that are yet to be incorporated into existing solutions. To determine them for the telemedicine apps in Bangladesh, we analyzed the reviews from Google Play and the survey responses.

**RQ3:** *What are the concerns of the user regarding the telemedicine apps in Bangladesh?* While using apps, users often raise different concerns, such as concerns about their security and privacy. We also studied such a perspective for the telemedicine apps in Bangladesh using Google Play evaluations and survey responses.

**RQ4:** *What are the challenges encountered by those responsible for managing and maintaining telemedicine apps? What do users think of them?*
Apps' sustainability relies not just on the apps' functionality but also on the people who maintain them. Therefore, we focus on this aspect for the telemedicine apps in Bangladesh to reflect the difficulties the managerial people confront and the users' thought processes about the managerial challenges.

**RQ5:** *Are user preferences and demographic traits correlated?* User preferences include both the functionalities of the apps and contemporary and future usages of technology. Therefore, we wanted to know whether they are anyhow related to the demographic traits of the users of the telemedicine apps of Bangladesh.

In a nutshell, our key objective in this study is to ascertain the perceptions of the people of Bangladesh on telemedicine apps. To do so, we have studied how the service recipients benefit from telemedicine apps, what features the service recipients like what challenges service providers to face from a managerial perspective, etc. We gathered user reviews from Google Play and conducted a qualitative analysis to accomplish so. Besides, we conducted a survey and assessed the responses quantitatively. We have also qualitatively analyzed feedback obtained from open-ended questions of the survey.

## 1.2 Our Contribution in This Study

We have made the following sets of contributions in this study:

(1) We collected user reviews on existing telemedicine apps available in Bangladesh. We collected the data in two steps and applied qualitative analysis to them.
(2) We surveyed the patients and collected 87 responses. We studied the opinions of the users who have used a telemedicine app their thoughts on the features of the telemedicine apps. Additionally, we examined the desired features of participants regardless of their prior experience with telemedicine apps. We performed $\chi^2$ test on the survey feedback and corresponding demographic traits.
(3) Based on our findings, we have explained the usability of telemedicine apps. We suggest ideal features and functionalities that a telemedicine app should provide and ways to overcome the barriers associated with telemedicine apps to provide health care services across Bangladesh efficiently.

## 2 RELATED WORK

Telemedicine is by no means a new concept. In broader terms, the history of telemedicine can be divided into three distinct eras: (1) the 1970s, which relied heavily on broadcast and television technologies; (2) the late 1980s and 1990s, which prospered from the digitization of telecommunications; and (3) the present era, which is dominated by internet-based telemedicine apps [17].

While reviewing the current literature on telemedicine apps, we uncovered research focusing on four distinct categories of work. Research on telemedicine apps globally, independent of a country's economic situation, research on telemedicine apps in Bangladesh. Research on the impact of telemedicine apps during the pandemic and research on the influence of telemedicine on specific specializations. Each of them has been discussed in further detail in the following subsections.

### 2.1 Telemedicine Apps Worldwide

With the advancement of technology, telemedicine apps became more sophisticated, and their use expanded. They are being used worldwide in developed, developing, and underdeveloped countries. In the United States, the majority of hospitals use some form of telemedicine to communicate with their patients [30]. Due to their technological advancements compared to developing and underdeveloped nations, developed countries have more significant potential to adopt telemedicine. They utilize telemedicine apps to narrow the gap between rural and urban health care services. On the other hand, while developing and underdeveloped countries increase their use of telemedicine apps, they confront different challenges. In developing countries, weak infrastructures are impeding the development of telemedicine apps [21]. Inadequate internet connectivity is likely one of the primary impediments to using telemedicine apps [2, 8]. Another barrier that is impeding the widespread use of telemedicine apps is the trust issue [2]. Bangladesh, being a developing country, is similarly confronted with these problems.

---
[1]https://play.google.com/store/apps

## 2.2 Telemedicine Apps in Bangladesh

Bangladesh is a densely populated country. Compared to the total population, the number of health care providers is negligible. Additionally, qualified doctors are generally based in cities, despite the bulk amount of people living in rural areas. Thus, telemedicine may be a solution to provide health care services while making the most use of a few resources [56]. Due to the large number of people who reside in rural areas [16], several studies have been undertaken on the rural health care system [37, 56, 62]. In 2009, the ministry of health introduced mobile phone health services in 418 Upazila health complexes and 64 district hospitals (a total of 482), where each had a mobile phone maintained by a doctor who would answer calls and give health care service over the phone [56]. According to research done by Khatun et al. [37], rural residents are less informed and ignorant of telehealth services. Even among those informed, telemedicine services were used by a minority. Often, individuals choose to obtain medicines from pharmacists or quacks. According to a study, these quacks should be integrated into government-supported infrastructure and subjected to strict supervision in order to deliver primary care [62]. Even if telemedicine apps are functional in rural locations, consumers who live close to cities would probably prefer in-person consultations [23, 48]. Apart from that, it has been demonstrated statistically that telemedicine app usage is dependent on age, literacy level, and gender [56]. Additionally, it is dependent on economic conditions too. Numerous studies have been undertaken on the use of telemedicine by underprivileged people [53, 65].

## 2.3 Rise of Telemedicine Apps during COVID-19

It has been observed that during the pandemic caused by COVID-19, the usage of telemedicine has increased significantly [36, 42] pointing to the need for study on the features offered by telemedicine apps. Not just only in Bangladesh, telemedicine utilization rose significantly all over the world with the outbreak of the COVID-19 [66] pandemic. To prevent the virus from spreading, social distancing was mandated, and towns were locked down for various periods of time. Personal visits to the doctors and hospitals became perilous. Telemedicine was vital during the pandemic, enabling some patients to get care from home and minimizing the infection rate [34]. This also enables elderly patients to consult through video conferencing, allowing them to remain at home [30]. Telemedicine apps enabled patients with diseases like diabetes to consult with a doctor from the comfort of their own homes, therefore preserving social distance [46]. People suffering from diseases like asthma and immunodeficiency are susceptible to COVID-19. Telemedicine has aided their treatment by allowing them less exposure to the world outside [52].

## 2.4 Use of Telemedicine Targeting Specific Specializations

Telemedicine can be adopted with a specific specialty of health care service in consideration. They can be pretty beneficial for monitoring maternal and newborn health [60]. There are opportunities to employ telemedicine apps in rural regions to ensure maternal health care [39] and to raise awareness about it [40]. Telemedicine has also been shown to be effective in the treatment of mental illness during the pandemic [58]. They have allowed patients to speak with therapists in the privacy of their own homes. Telemedicine enables patients who seek care for anxiety or depression to receive assistance without having to visit a hospital, and psychological stabilization therapy may be offered via the app [27]. Telemedicine has many benefits for allergy and immunology services, including the reduction of health care providers' exposure to potentially infected patients [46]. Additionally, a study demonstrated that telemedicine apps could be beneficial in managing heart failure and diabetes [24]. Furthermore, telemedicine is not confined to doctor-patient interaction; it also encompasses the integration of health care networks, allowing clinicians to communicate with one another [3, 9].

## 3 METHODOLOGY

The previous section's aggregated findings from the literature review reveal that telemedicine apps have been evolving. They are gradually gaining acceptance around the world. The COVID-19 epidemic, in particular, has fueled the growth of telemedicine apps. They are even being designed to focus on specific specializations. Telemedicine has been studied in the setting of Bangladesh. However, we noticed a lack of research on the features and functionalities offered by telemedicine apps in Bangladesh. To better understand peoples' perspectives towards them, we collected publicly available reviews of several telemedicine apps. We also conducted a survey focusing on the patients.

### 3.1 Data Collection from Google Play

In Bangladesh, different mobile operating systems are available. Among all the smartphone users, the Android users are relatively large [12]. This is why we chose Google Play for collecting user reviews. Users can rate and review apps on Google Play. We wrote a script to collect these reviews. We selected 14 telemedicine apps for review based on their rating and number of downloads. We utilized three keywords in our search for telemedicine apps, *telemedicine Bangladesh*, *mhealth Bangladesh*, and *telehealth Bangladesh*. Not only did the script get the reviews, but it also provided important insights about the apps. With each review, there is a rating incorporated. The rating is based on a 1 to 5 point scale. We gathered data in two stages. We began by collecting a total of 20 reviews for each possible rating of the apps, i.e., 20 reviews that have been rated 1-star, 20 reviews that have been rated 2-star, and so on. Detailed ratings of a few apps had less than 20 reviews. As a consequence, several apps received fewer than 100 comments. We got over 1100 reviews in total from 14 apps. Then we used thematic analysis [19, 20] to evaluate the 1100 reviews and extract themes from them. Following that, we ran the scraper once again. We gathered 100 reviews for each rating, totaling 500 for each app. Since some of the applications had fewer ratings, we gathered a total of 3800 reviews. Then we applied thematic analysis once again. Each of the reviews has been assigned a unique id. The naming format is as follows: *A*, followed by an app id ranging from 1 to 14, *S*, and lastly, the review id, e.g., *A1P2* means Participant 2 of app 1. Throughout the article, we have followed this convention whenever we have quoted a participant from the reviews. Table. 1 contains a list of the selected apps.

### 3.2 Survey

While the reviews are not just limited to the app's features, we have observed that most of the reviews are incorporated with the features. That is why we have surveyed to get more insights into the features and functionalities of telemedicine apps.

*3.2.1 Recruitment.* We gathered information using two methods: personal relationships of the authors and snowball sampling [18]. We communicated with the persons through email and social media. We contacted the connections and requested to take part in the survey. Additionally, we asked them to forward it to their friends and relatives. Furthermore, We posted our survey links which were taken using Google Forms[2] to multiple Facebook[3] groups. The survey was voluntary, and no compensation was given.

*3.2.2 Participants.* We provide a highlight of our participants' demographic traits in Table. 2. We prepared a questionnaire for the patients. We managed to collect the opinions of 87 patients. Among 87 patients, a total of 11 have used telemedicine apps. Our participants vary in age, gender, literacy level, and where they live. We have addressed these participants with a unique id throughout this paper with a prefix *SP*.

*3.2.3 Procedure.* We surveyed the general public. For the users, the questionnaire was set both in English and Bengali. Each questionnaire took between 15 and 20 minutes to complete. To begin, we gathered demographic data. Then we asked questions about their working hours and their comfort level with electronic devices. We solicited input on already available telemedicine apps from users who have used telemedicine apps. Finally, we seek feedback on the desired features and functionalities from all. For the opinion on features and functionalities, we used the Likert scale [33]. The Likert scale had five options, (1) strongly agree, (2) agree, (3) neutral, (4) disagree, (5) strongly disagree.

### 3.3 Analysis

We performed thematic analysis to analyze the data we gathered from Google Play and some open-ended survey questions. Additionally, we saw what most users desire from telemedicine apps based on survey data. On the reviews, we utilized open coding [28]. Initially, the first author reviewed the reviews collected on the first run and developed codes. Following that, all contributors revised the codes and developed themes. Finally, after collecting the reviews for the second time, the authors reviewed them one more time to ensure that nothing was overlooked.

For the survey responses, we performed the $\chi^2$ test [51] with a significance level of 0.05. We have tried to incorporate eight user preference-specific questions with four demographic traits and check whether they can be correlated or not. As the questions are large, we shall address them using abbreviations in this paper. Table 3 contains the list of the abbreviations and Table 4 contains the result of our analysis. While performing the $\chi^2$ test, we excluded those groups from demographic traits with no data, e.g., no data was found for the age group 41-45 in Table. 2, so we omitted that. Also, we excluded those Likert chart values for which there was no response for a particular user preference specific question, e.g., no one responded *Neutral* for a particular user preference specific question.

### 3.4 Ethical Concerns and Approval

The data collection of this study was applied to the Ethics Committee of the corresponding author of this paper and got approved accordingly. Besides, the authors complied with ethical concerns of the data by maintaining the security and privacy of the data.

## 4 FINDINGS

Our assessment of Google Play user reviews reveals a plethora of information. We have observed that there is both positive and negative feedback for each app. Users who benefited wrote positive things; some of them even thanked them for taking such initiatives. On the other hand, users who have faced issues while using the app reported bugs, criticized various features, and raised concerns about privacy and security. Furthermore, users have expressed desirable features. Our survey questionnaire also focuses on different features of telemedicine apps, giving us more insights into them. The following sections discuss users' perceptions of telemedicine apps, including what users praised, what users complained about, what users desired, and how users got positively impacted.

### 4.1 User Perceptions of Existing Telemedicine App Features (RQ1)

The majority of the telemedicine apps have some key features. They provide video consultations with doctors. The app includes a list of doctors. The user can see their details, including their schedule. Appointments can be scheduled, rescheduled, and canceled. Certain telemedicine apps provide alerts prior to the appointment. After the consultation is complete, the doctor electronically provides the user with the prescription. The majority of telemedicine apps collect consultation fees digitally via Mobile Financial Systems [13], credit or debit cards, etc. Some apps provide scheduling of diagnostic tests physically in the diagnostic centers. Some apps work as an online medicine store where the users can order and get home delivery of medicines. Among the applications we examined, one includes a forum for users to ask questions to specialists in addition to the video consultation option. Among all the features, some are really appreciated by the users; some are criticized. Moreover, they have suggested new features according to their need. They have reported bugs they have frequently encountered and expressed their overall

---
[2]https://www.google.com/forms/about/
[3]https://www.facebook.com/

| App name | Release date | Total downloads | Average rating | Total rated by | Total reviews |
|---|---|---|---|---|---|
| Doctime | May 13, 2020 | 500,000+ | 4.65 | 6780 | 2336 |
| Sebaghar: Online Doctor Video Consultation App BD | May 30, 2019 | 100,000+ | 4.23 | 1452 | 518 |
| Doctor Dekhao | April 1, 2020 | 50,000+ | 4.05 | 393 | 151 |
| LifePlus Bangladesh | Jan 27, 2022 | 10,000+ | 4.8 | 415 | 93 |
| Maya - It's ok to ask for help | Feb 2, 2015 | 1,000,000+ | 4.3 | 14646 | 5596 |
| Arogga - Online Pharmacy and Healthcare App | Feb 12, 2020 | 100,000+ | 4.22 | 1399 | 699 |
| MetLife 360Health Bangladesh | Nov 18, 2021 | 10,000+ | 4.51 | 84 | 44 |
| MedEasy - Online Pharmacy BD | May 8, 2020 | 10,000+ | 4.7 | 309 | 232 |
| Daktarbhai - Your health companion | Sep 21, 2017 | 100,000+ | 4.3 | 1566 | 643 |
| Digital Hospital | Mar 21, 2020 | 100,000+ | 4.11 | 1191 | 563 |
| Patient Aid | Apr 28, 2016 | 1,000,000+ | 4.5 | 36989 | 9586 |
| DIMS | Nov 24, 2014 | 1,000,000+ | 4.57 | 37314 | 13943 |
| My Health - Health and Doctor Consultations | Nov 27, 2018 | 50,000+ | 4.23 | 709 | 337 |
| Tonic | Dec 21, 2016 | 100,000+ | 3.9 | 1495 | 516 |

*All the metrics including total installs, rating, total rated by, and total reviews were retrieved on Jan 31, 2022*

**Table 1: Information of telemedicine apps that has been analyzed.**

user experiences. This research question has mainly been answered by analyzing the reviews from Google Play.

*4.1.1 User Favored Characteristics.* Users who have benefited from telemedicine apps rarely mentioned specific features they loved. Several people stated that the list of doctors was beneficial to them as they could easily find out doctors according to their problems. Others praised the app as a whole. A1P293, who is a doctor, mentioned, *"Good app for patients and doctors. I get immense pleasure when I give treatment sitting at home to patients situated at their homes at a distance—many hearty congratulations to Doctime authority and customers for making telemedicine grow faster in the country"*. The majority of users expressed satisfaction with the doctors they consulted. A1P298 expressed their sentiments, *"The doctor is incredible. Not only has she taken great care of my health, but also she is lovely to speak with at every appointment. It's rare to find a doctor that combines such personal touches and care for a patient as a person with an outstanding quality of medical care. I highly recommend becoming her patient!"*.

The question-and-answer platform implemented by *Maya- it's ok to seek assistance* has been exceptionally well received by users. Users have stated their reluctance to share and discuss some topics with relatives and friends. This forum enables people to discuss those issues anonymously. A5P487 mentioned that, *"Sometimes we are not able to share our personal experiences with our parents and we like to share with our same age friends but this should not be the case. Luckily with the help of Maya, I could easily reach trained doctors and clarify my queries"*. Additionally, the app provides health-related articles, which have been well-received by users.

Apps that offer home delivery of medications have been widely praised. Home delivery of goods is becoming popular in Bangladesh. Medicine is no exception. Obtaining medication online and having it delivered to their doorsteps at a reasonable price has benefited people, particularly during the epidemic.

*4.1.2 Frequently Reported Bugs.* Users have reported bugs they have encountered in their reviews. The majority of them were discovered to have authentication issues. When mobile numbers are verified, the apps send an OTP (One Time Password) to validate the user. Users have reported that the OTP does not work correctly. A1P31 mentioned, *"OTP was not being sent to my number. I tried to resend it a few times but it's not working"*.

There have been server issues during critical times. A1P9 said, *"Didn't get the prescription in an emergency situation due to server problem"*.

Numerous users have had trouble while paying. These are bugs related to the Mobile Financial System payment gateway API [22]. Users have reported several occurrences of unsuccessful transactions. Apart from this, a few users reported having difficulty with video calls, scheduling appointments, etc. It was discovered that most users faced issues when the apps got new updates.

*4.1.3 Ease and Comfort of Use.* We discovered diverse opinions about the UI/UX of the apps we examined. Users that expressed dissatisfaction with the UI/UX said that the icon and typography are improper. A1P56 stated that, *"The initiative is great but the font is tiny and the icons are not intuitive"*. Additionally, they have expressed dissatisfaction with the performance of the app. Certain users became stranded on the loading screen, significantly degrading the user experience. A12P21, a doctor, stated that after using the app for a couple of years, they were forced to transfer to another telemedicine app due to a poor user experience: *"I am a doctor and I am using this app for 2-3 years. It's beneficial whenever I wanted to choose drugs from different companies. But I have had a bad experience due to loading when I have to open this app. At first, it shows*

| Gender | Participants percentage |
|---|---|
| Female | 20.7% |
| Male | 78.2% |
| Prefer not to say | 1.1% |
| **Living Aspects** | |
| Inside the capital | 71.3% |
| Outside the capital | 28.7% |
| **Age-range** | |
| 16-20 | 20.7% |
| 21-25 | 51.7% |
| 26-30 | 20.7% |
| 31-35 | 3.4% |
| 36-40 | 0% |
| 41-45 | 0% |
| 46-50 | 0% |
| 51-55 | 2.3% |
| 56-60 | 0% |
| 61-65 | 1.1% |
| 65+ | 0% |
| **Literacy level** | |
| Below High School/SSC/O-level/Equivalent Degree | 0% |
| High School/SSC/O-level/Equivalent Degree | 5.7% |
| College/HSC/A-level/Equivalent Degree | 26.4% |
| Above College/Diploma/Equivalent Degree | 5.7% |
| University/Higher Degree | 62.1% |

**Table 2: Summary of patients' demographic traits.**

*loading animation, then makes my phone hang or doesn't respond. If I switch off my data connection instantly it becomes active from frozen. Now I have switched to bdf 360 app for this problem"*. Users have reported that certain apps show advertisements that irritate them. They said that they were confronted with advertisements each time they looked for medications.

Individuals of all faiths can utilize telemedicine apps. We have noticed that religious views affect the UI/UX. A12P58, a devout user, expressed worry over possible violations of his Purdah. Purdah is a term that refers to a 'curtain' or a covering worn by Muslims [47, 50]. He explained, *"Helpful App for Doctors but the app's user interface should be changed. A woman's picture is shown on the homepage of the app for which Purdah is difficult for male users"*.

Bangladeshi people speak Bengali as their first language. The apps should be offered in both Bengali and English, as users require. The user requests that the program support two languages. Numerous websites and programs allow for theme customization. Certain users requested a dark theme to enhance the user experience.

## 4.2 Features Desired by the Users in Telemedicine Apps (RQ2)

Users have proposed features they would like to see implemented. This research question has been answered by analyzing the reviews from Google Play and the responses from the surveys.

Among the most frequent of the desired features is communicating with the doctor through chat. The users wish to communicate with the doctor via text messages. In modern messaging apps such as Messenger[4], WhatsApp[5] the active status of the user can be seen. We have found that most users are interested in the doctor's active status and have no trouble informing doctors of their availability. Those who did not wish to disclose their active status expressed privacy concerns. SP32 said, *"Some people wouldn't prefer letting doctors inform about their online status because of privacy concerns"*.

As previously stated, most applications collect fees via Mobile Financial Services. Additionally, other users stated that they paid with credit cards. Not all applications include all of Bangladesh's popular Mobile Financial Services. As a result, people desire to see all popular Mobile Financial Services integrated. A1P217 mentioned, *"It would be great if they incorporate payment through Rocket and Nagad along with bKash"*.

When users log into the app, they do so as a single individual, and only that individual's medical information and history are kept. On the other hand, the users desire something for their families—a profile system to store medical information and family history for all family members. Specific applications have been found to offer family subscriptions.

As with any other service, the user wants to rate doctors. They wish to rate and comment on the doctor. SP86, on the other hand, raised concerns about rating using the Likert scale [33] or similar systems. They stated, *"Be careful with any ranking/scoring mechanisms like five stars. It might not be the best way to judge doctors and can be misleading to prospective patients"*.

Some users requested a scanner with a handwriting recognizer to scan and upload their old prescriptions. Furthermore, users desire the ability to upload reports. Users of apps that solely offer video consultations expressed a desire for an online medicine delivery system. Some users seemed to be confused with navigation inside the app. S79 suggested that, *"It would be great to have a video tutorial after we install the application"*.

## 4.3 Privacy and Security Concerns of the Users Regarding Telemedicine Apps (RQ3)

Like any other app, telemedicine apps also have privacy and security issues. They deal with a lot of patients' data that should be only accessible by the patient and the doctor they are consulting. Moreover, these apps use third parties for handling payment where the users have raised their concerns. Finally, these apps seek many permissions in smartphones, some of which seem redundant to the users. This research question has mainly been answered by analyzing the reviews from Google Play and partly by the responses from the survey.

---
[4]https://www.messenger.com/
[5]https://www.whatsapp.com/

| Abbreviation | User preferences |
|---|---|
| Usg | I feel comfortable using smartphone or computer |
| Comm | I feel comfortable in communicating with people, e.g., relatives, friends, etc., through Zoom, Google Meet, WhatsApp, or Facebook |
| Cslt | I would prefer to consult with the doctor through Zoom, Google Meet, WhatsApp, or Facebook |
| Trtd | There are some cases in which patients can be treated through telemedicine apps |
| Dstat | I would like to see the online status (whether the doctor is online or offline) of the doctor with whom I have an appointment |
| Mstat | I would like to let the doctor know my online status (whether I am online or offline) |
| Ntfn | I would like to get a notification before my appointment starts |
| Pmnt | I would like to pay for the appointment immediately after finishing consultation as per the appointment |

Table 3: Abbreviated terms for different user preferences from the questionnaire.

*4.3.1 Trust Issue with Payment.* In Bangladesh, Mobile Financial Services (MFS) are increasingly becoming popular [54]. The majority of telemedicine apps collect consultation fees through Mobile Financial Services. A few offer cash on delivery for their medicine home delivery service. Some of the well-known Mobile Financial Services in Bangladesh [64] are bKash[6], Rocket[7], Nagad[8], etc. Payment API Gateways from these mobile financial services are available to integrate telemedicine apps. When the user reaches the payment screen, the app requests their Mobile Financial Service credentials to process the payment. This is where some users have expressed their concerns. Some users reported that the app requested their pin, which nobody shared with them. A5P22 mentioned, *"The answers provided by the experts are very useful. So, I wanted to purchase a package, but during the process, they asked me for my bKash pin. I tried to do it in another way but providing the pin was the only way. Why would people give you their bKash pin?"*. Some users recommended that payment should be handled independently by Mobile Financial Service apps and that telemedicine apps could request the Mobile Financial Service app's transaction id to validate the payment. A1P14 said, *"Why would they ask for my bKash information? They should provide a number where I can transfer the money and provide a box in the UI to insert the transaction id"*.

*4.3.2 Concerns about Unnecessary Permissions.* In Bangladesh, privacy leakage via apps has become a significant issue [5]. Like any app, telemedicine apps require specific access permission of a smartphone in order to function correctly. They require camera access for video calling, microphone access for voice calls, and so on. Certain users have expressed that certain apps are requesting unnecessary permissions. For instance, a telemedicine application solely delivers information on medications requires storage and media access. The concern of the users is visible in the comment of A11P36, *"An app developed only to retrieve medicine information does not need personal storage access. They are stealing people's personal data."*. According to some users, the apps are stealing personal data by granting unnecessary permissions. Users believe that an app should be granted permissions when it is required.

---
[6]https://www.bkash.com/
[7]https://www.dutchbanglabank.com/rocket/rocket.html
[8]https://nagad.com.bd/

## 4.4 Challenges from a Managerial Perspective (RQ4)

When evaluating the users' experience, we should not confine our analysis to the features and functionalities of the app. The organization that manages the app significantly influences the users' perception. They act as a bridge between doctors and patients. Their role is to manage various aspects of the app, such as customer support, responding to user complaints, and incorporating doctors according to the users' needs. The reviews from Google Play have answered this research question.

*4.4.1 Communication with Customer Service Representatives.* Users generally contact the customer care service when they face any issues. Customer care service is not available in all the selected apps for our study. Users have felt the need for customer care service for those apps. While there is positive feedback on the customer care service, there is much negative feedback as per our observation Some of the users' have complained that the customer care service is non-cooperative. Even there is a case where the user got blocked by customer care on Whatsapp while complaining and demanding a refund. In this regard, A3P1 commented, *"I had an appointment with the doctor at 7:40 PM, but the call didn't connect and I tried several times. Contacted the helpline and they said that they will reschedule my appointment at 11 PM but after that no response from the team. I was in a serious need to talk to the doctor but I could not. I messaged them on WhatsApp and told them to refund me but they blocked me on WhatsApp."*. A1P12 mentioned, *"Irresponsible customer support, contacted 4 times for an urgent issue, but within 1 hour they didn't come up with any solution. When it's possible to solve this within 10 minutes"*.

*4.4.2 Issues Regarding Consultancy Fee.* In case of in person consultations, different doctors have different consultation fees. Some telemedicine apps charge a fixed price for consultations, while others charge a fee that varies by doctor. There are premium packages for some apps. Some offer monthly packages too. We saw several apps raise their fees, which users didn't appreciate. Additionally, users recommended introducing a follow-up fee. SP59 proposed that, *"If the patient requires a physical examination following the telemedicine session, the doctor may consider decreasing their regular fee"*.

*4.4.3 Managing Doctors According to Expertise.* Individuals might be affected by a variety of ailments. Thus they need doctors of different expertise. We have observed that there is a lack of expert doctors on some platforms. From the reviews, we saw that people demanded doctors from various specializations. Additionally, we have identified a demand for emergency doctors. In general, the patients have to check out the doctors' schedule, set an appointment, and then consult. However, it is logical to seek advice from a doctor in an emergency. As a result, users emphasized the importance of doctors. While telemedicine apps are intended for people, one user recommended including veterinary doctors for pets. A2P107 suggested *"I downloaded it for consulting a pet doctor but there was no option for it. I hope they add it soon"*.

## 4.5 Correlation between Demographic Traits and User preferences (RQ5)

We have taken eight user preference-related questions and four demographic traits from our survey to check whether they are correlated. For this, we have conducted $\chi^2$ testing where the significance level is 0.05. We have mapped our 8 user preferences in Table 3. We have found out that 31 out of 32 tests got accepted as the calculated value is less than the $\chi^2$ distribution table value [11]. The only one test that got rejected is between age and the statement that there can be some cases where treatment can be provided through telemedicine apps. In Table 4 we have presented the results found.

## 5 DISCUSSION

In this section, we shall discuss the implications of our findings and make recommendations on what a telemedicine app should provide to users to be successful in Bangladesh.

## 5.1 Features a Telemedicine App should Provide

The majority of telemedicine apps have a similar set of features. The user creates an account, looks for a doctor who meets their criteria, sets an appointment, and receives a consultation on that day through video or voice calls. Apart from these fundamental features, some apps provide additional functionalities, such as ambulance service, diagnostic test booking, etc. According to our observations, users sought a variety of features from telemedicine apps. Thus, based on our results, we propose certain features to please the users.

*5.1.1 Online Chat Option.* The ability to send a message to the doctor is one of the most sought features by users. There may be situations when the patient forgets to ask the doctor a question during the consultation. They might ask it later via text messages. The same is true for the doctor. However, we would suggest limiting the ability to send messages. The texting option can be limited to a specific period before and after an appointment, for example, half an hour before and after the appointment. Otherwise, patients may send text messages to doctors on an ongoing basis without an appointment. Along with this functionality, the app should allow both the user and doctor to see each other's online active status. This option, however, should be optional due to the privacy concerns of both the user and the doctor.

*5.1.2 Easy Rescheduling and Cancellation of Appointments.* Both the patients and the doctor may need to cancel or reschedule the appointments. If the patient wishes to cancel the appointment, they should be refunded. If they choose to reschedule, the appointment should be moved to a time convenient to both of them. In the rescheduling process, the main concern is with the payment. If the consultation fee is collected immediately after the appointment is scheduled, the patient must be refunded in the event of cancellation. We propose that the organization that manages the app act as a payment intermediary. The patient would pay the organization after scheduling an appointment, and the organization would pay the doctor after the consultation is finished.

*5.1.3 Integrating Popular Mobile Financial Services.* In Bangladesh, a few Mobile Financial Services (MFS) have gained considerable popularity, including 'bKash,' 'Rocket,' and 'Nagad.' Because a user may utilize any of them, the leading Mobile Financial Services (MFS) payment gateways should be integrated with telemedicine apps. Additionally, payment through debit and credit cards should be incorporated.

*5.1.4 Giving Rating and Feedback of Doctors.* The patients should rate the doctors and write reviews about them. When a patient seeks doctors to meet their specific needs, they may use the app to search by specialization. Multiple results can be found when searching for a doctor. This is when the rating and reviews play a critical part in determining which doctor to choose. According to our results, one user expressed worry about ranking doctors on a 5-point scale since it may become biased. We would recommend asking a few questions after the consultation is over, from which the rating will be determined. These questions will cover the doctors' timeliness, demeanor, and how happy the users are with the doctors' advice.

*5.1.5 User Profile for Multiple Persons.* In order to use a telemedicine app, at first, a user must register. They acquire a profile after registering. Additionally, the app stores their medical history. However, a user may utilize a single app for all family members. In such a situation, the app should allow for the addition of numerous members and the maintenance of distinct medical histories for everyone. The majority of available apps do not support multiple users on the same device [5]. However, telemedicine apps must provide multi-user access.

*5.1.6 UI/UX Enhancement.* When building an app, a user-centered design [1] approach should be adopted. People of different ages can use these apps. Not everyone, regardless of age, has the same intuition when utilizing a smartphone or an app. Elderly individuals frequently require assistance with these apps [38]. As a result, telemedicine apps must be intuitive. They should be customizable in terms of fonts and color schemes. According to our observations, some people like light themes, while others prefer dark. As a result, the app should have choices for altering the theme. Additionally, the icons should be intuitive.

Dual language support should be included in the app. Bengali is the native language of Bangladeshis. Many users may feel uncomfortable using the app if it is only English. Thus, an app needs support in both Bengali and English.

| User preferences | Demographic traits | | | |
| --- | --- | --- | --- | --- |
| | Age | Gender | Living aspects | Literacy level |
| Usg | Accepted | Accepted | Accepted | Accepted |
| Comm | Accepted | Accepted | Accepted | Accepted |
| Cslt | Accepted | Accepted | Accepted | Accepted |
| Trtd | Rejected | Accepted | Accepted | Accepted |
| Dstat | Accepted | Accepted | Accepted | Accepted |
| Mstat | Accepted | Accepted | Accepted | Accepted |
| Ntfn | Accepted | Accepted | Accepted | Accepted |
| Pmnt | Accepted | Accepted | Accepted | Accepted |

Table 4: Result of the $\chi^2$ testing.

## 5.2 Protecting Users Privacy and Security

With the increasing usage of apps, the security issue is becoming alarming. Telemedicine apps, in particular, must take great care to protect users' privacy, as health data is deemed sensitive [26, 63]. We discovered three ways telemedicine apps potentially infringe on users' privacy and security, which we will briefly discuss.

*5.2.1 Privacy of Users Medical Records.* Telemedicine offers the ability to store people's health records digitally. This would make it simple to access them at any time. There is a risk of losing medical records with the conventional method of maintaining physical copies of reports and prescriptions. However, with telemedicine, data may be saved on the cloud with minimal risk of loss. These are private and sensitive data available to only patients and healthcare providers. Numerous countries have enacted legislation to this effect [26, 59]. In Bangladesh, privacy leakage via applications has become a severe worry [5]. Thus, the government should monitor telemedicine apps for data breaches. Since the consultation is conducted through video call and any patient may discuss their intimate areas, it is vital to check that the videos are not recorded.

*5.2.2 Ensuring Secured Payment System.* According to our findings, most consultation fees are paid via Mobile Financial Services. Some users expressed concern about this, as the apps require the users' MFS account PIN for payment. They mentioned that they would feel safe paying using the MFS apps and provide the transaction id after the payment is made. This manual approach, in our opinion, would be inefficient since it would require human intervention to verify transactions. We propose that telemedicine apps utilize a payment gateway such as Uber[9] that has incorporated 'bKash' into their system for Bangladeshi customers, requiring them to connect their 'bKash' account to their 'Uber' account once by providing their PIN. The connection of the two accounts is secure since it is authenticated using OTP. The user gets alerted via the 'bKash' app whenever a transaction is done. Thus, even if an unauthorized transaction is made, the user will detect it immediately and take appropriate action.

*5.2.3 Restricting Various Permissions in the Smartphone.* Like any other app, Telemedicine apps require permissions to access certain functionalities on a smartphone. The camera and microphone are the most typical items to which apps require access. However, we discovered that some apps had requested permission to access unnecessary features for the app to function correctly. As a result, several users have expressed concern regarding these apps. In Bangladesh, stealing user data without their consent is nothing new. The popular ride-sharing app 'Pathao' was accused in 2018 of collecting data from users that were not essential for the app to function effectively [41, 43, 55]. The majority of app users are unaware that their data is being leaked. So, it is the government's responsibility to keep an eye on these apps to see whether these apps are violating laws by stealing user data without their consent.

## 5.3 Effectiveness of Telemedicine Apps

Telemedicine services have proved themselves to be quite beneficial. Users got benefited in numerous ways, notably throughout the continuing epidemic. Users have found the doctors to be professional most of the time. The app also served as a tool to find the appropriate doctor. It has saved their time and reduced the need to commute and visit doctors in person.

*5.3.1 Saving Time and Commuting.* Telemedicine apps are saving users time to a great extent. Roads in the cities of Bangladesh stay congested for the majority of the time. Thus, visiting doctors in person can be time-consuming. However, adopting a telemedicine app eliminates travel time, saving patients valuable time. Telemedicine apps have also proved effective for patients who find it challenging to move physically. Patients frequently go to the capital city for physical examinations. After the doctors have completed their treatment, they ask the patients to follow up on specific days. These can be accomplished through the use of telemedicine apps.

*5.3.2 Allowing Discussion of Topics People are Hesitant to Share.* Individuals who are hesitant to ask questions directly to doctors have found these forums quite beneficial. Individuals may be hesitant to discuss their concerns about their private parts or something about their mental health. Telemedicine apps may be an option for them. Among the apps we have studied, *Maya-it's ok to ask for help*[10] allows users to ask questions anonymously. In this regard, A5P22 stated, *"It's a wonderful app. It's so helpful for those people who do not have a person to share his personal matter or private physical problem. This app will be your psychologist and your personal doctor"*.

---

[9]https://www.uber.com/bd/en/

[10]https://play.google.com/store/apps/details?id=com.maya.mayaapaapp

*5.3.3 Availing Doctor Consultation During Pandemic.* The ongoing COVID-19 pandemic has resulted in a significant change in consultation with a doctor. Typically, our people have visited doctors in person. However, it is pretty risky to visit the doctors in person due to the possibility of contracting with COVID-19. Additionally, doctors are in danger. There have been cases where the patient did not disclose that they were COVID-19 positive and affected the doctor. Amidst the pandemic when the healthcare system is in crisis in Bangladesh [6], telemedicine apps have allowed people to consult with doctors in a safe environment.

*5.3.4 Finding the Right Doctor.* Telemedicine apps have facilitated finding the right doctor. The apps provide available doctors with their information and expertise. Users may pick doctors according to their requirements and fix appointments. Some apps suggest doctors nearby go to the nearest doctor if they need in-person consultancy. Users also reported that they had found telemedicine apps beneficial at times of emergency. A14P287 mentioned, *"My father was having pains in his chest. Tonic provided me instant doctor support and calmed us down by saying it was just indigestion. The doctor also prescribed the medicines which subsided it"*

## 5.4 Barriers to Telemedicine Apps

Despite their usefulness, telemedicine apps are not widely used. Specific reasons are refraining people from using these apps and availing healthcare services. Both technological and non-technological difficulties constitute an impediment, which we will discuss briefly.

*5.4.1 Digital Literacy.* One of the reasons for the lack of utilization of telemedicine apps is a lack of digital literacy [25]. This can vary because of age and educational qualifications. Individuals with a higher level of education are more likely to use telemedicine apps [37]. On the other hand, those with a lower level of education are less aware of telemedicine services [37]. Older adults are usually hesitant to use electronic devices. Thus, age plays a vital role here. They are reluctant to new technology and unwilling to adapt to new technology. However, this can be overcome by taking help from a technologically educated junior family member. Another way to overcome this barrier is to involve a simplistic design principle that helps overcome digital illiteracy.

*5.4.2 Poor Condition of the Internet.* A strong Internet connection is required to make successful video and audio calls while using telemedicine apps. Like many developing countries [2] the Internet speed in Bangladesh is not up to the mark [29, 31]. Additionally, Internet speed varies by location. Rural areas lack broadband and rely heavily on 2G or 3G cellular networks [62]. Thus, a slow Internet connection limits the widespread use of telemedicine apps. The doctor needs to know various test results and symptoms to make a proper diagnosis. It is not easy to share if the internet speed is not adequate. Thus, it is hard to fulfill the goal of a telemedicine application.

*5.4.3 Tradition and Lack of Trust.* Traditionally, the people of Bangladesh are not that much interested in going to a hospital or visiting a doctor. Sometimes, they consult with a local pharmacist or alternative medical service provider to have their medical need. It is a common tradition in rural areas and the country's cities.

Telemedicine apps can provide a one-stop solution by replacing this traditional treatment with no scientific background. Another thing is that they lack trust in having medical treatment. They usually trust the people based on their familiarity or referral, not just the qualification of the healthcare provider. With the proper advertisement of the telemedicine apps, people can know more about their benefits and may grow trust in them.

## 6 LIMITATIONS AND CONCLUSION

The research questions have mainly been answered by analyzing the reviews from Google Play and partly by the responses from the surveys. We evaluated the Google Play reviews that were scraped. While we obtained some fascinating facts regarding users' impressions of telemedicine apps, we could not correlate them to the demographic information of the users who gave a review. The data we retrieved contains no information about the users' demographics. As a result, we discovered no correlation between the features and demographic information provided by individuals in their reviews. Additionally, we believe that if we could read all of the comments on all of the accessible telemedicine apps, we would discover additional exciting details that we could have overlooked in our current approach.

We conducted statistical analysis on the survey responses to ascertain relationships between demographic data and characteristics. Additionally, we attempted to determine what most users desire based on the percentage of chosen responses. Due to the small number of participants and the fact that we employed snowball sampling, the demographic information is less diverse. The majority of respondents to our study are under the age of 30. The majority of them dwell in Bangladesh's capital city. As a result, the survey results cannot be used to derive generalizable conclusions. To make it more concrete, we need people of all ages and those who live in rural regions of Bangladesh.

This work focuses on the patients' thoughts. Though we came across a few doctor reviews, we could not correctly extract what doctors believe about telemedicine apps. Doctors in Bangladesh are extremely busy, and because the doctor-patient ratio is so low [7, 14], it was difficult to interview them. However, we are considering this as a future study in which we will extract the opinions of not just doctors but of all types of persons involved in the healthcare system.

We learned a great deal about users' attitudes regarding telemedicine apps despite all the restrictions. The evaluations on Google Play, in particular, provided valuable insight into the issues and desires of users. Additionally, we found their privacy problems, which is critical. We have discussed the usefulness of telemedicine applications, the obstacles that stand in their way, and how to overcome them. Additionally, we made recommendations for a telemedicine app's features. We promote more research in this sector, with a particular emphasis on maternal health. We also promote research into how telemedicine apps might assist with family planning and other essential topics like menstruation and sex education, considered taboo in Bangladesh.